\documentclass{PoS}

\title{Symbolic integration and multiple polylogarithms}

\ShortTitle{Symbolic integration and multiple polylogarithms}

\author{\speaker{Christian~Bogner}$^a$ and Francis~Brown$^{ab}$ \thanks{We thank Humboldt University for hospitality and support. FB is partially supported by ERC grant no.\ 257638.} \\ \\
       \llap{$^a$} Institutes of Physics and Mathematics, Humboldt-Universit\"at zu Berlin\\
        Unter den Linden 6, 10099 Berlin, Germany\\ \\
       \llap{$^b$}  Institut des Hautes \'Etudes Scientifiques \\
       Le Bois-Marie 35, route de Chartres, 91440 Bures-sur-Yvette,  France\\

        E-mail: \email{bogner@math.hu-berlin.de}, \email{brown@math.jussieu.fr}}

\abstract{ We review  a method for the algebraic treatment of  a family of functions which  contains  the multiple polylogarithms, with applications to the symbolic calculation of Feynman integrals.}

\FullConference{Loops and Legs in Quantum Field Theory - 11th DESY Workshop on Elementary Particle Physics,\\
		April 15-20, 2012\\
		Wernigerode, Germany}

\begin{document}

\section{Introduction}

Over the decades, polylogarithms have gained importance in perturbative Quantum Field Theory ever since the first occurrences of the 
dilogarithm in early QED (e.g.\ \cite{Rac}) and in results for one-loop integrals. As the complexity of the computations has increased, the literature on Feynman integrals has gradually absorbed 
a variety of generalizations, including Nielsen and  classical polylogarithms \cite{Lew, Nie, KMR}, harmonic polylogarithms \cite{RemVer} and later  generalizations \cite{GehRem, AglBon, BGM04, AblBluSch},
(some of which were previously known to 
 mathematicians by the name of hyperlogarithms \cite{Lap}), and variations on  multiple polylogarithms \cite{BBBL, Gon01}. For several approaches to computing Feynman integrals, it is useful to represent these
functions as iterated integrals.

In this talk we discuss a class of functions which was studied in reference \cite{Bro06}. These functions are closely related to (and contain) the multiple polylogarithms of Goncharov \cite{Gon01} and admit special properties which are useful
for the computation of Feynman integrals. In section \ref{sec: homotopy} we briefly review an important result by Chen \cite{Che} on the conditions for an iterated integral to give 
a well-defined function of several variables. In section \ref{sec: polylogs} we define a map whose image satisfies these properties and use it for the 
construction of the mentioned class of functions. Section \ref{sec: Feynman} briefly shows how these functions can be used in a systematic approach for integrating over 
Feynman parameters, presented in reference \cite{Bro08}.  Algorithms and a computer program for using this class of functions will be
the content of another publication, which is currently in preparation.

\section{Homotopy invariance and Chen's theorem \label{sec: homotopy}}

The property of homotopy invariance is best discussed when viewing iterated integrals as integrals along paths. Let ${k}$ be the field of either the real or complex numbers and
${M}$ a smooth manifold over ${k}$. Let a piecewise smooth path ${\gamma}$ on ${M}$ be given by a map ${\gamma : [0,\,1]\rightarrow M.}$ Two such paths ${\gamma_1}$, ${\gamma_2}$
are said to be \emph{homotopic} if their endpoints coincide:  ${\gamma_1(0)=\gamma_2(0)=x_0}$ and ${\gamma_1(1)=\gamma_2(1)=x_1}$, and if furthermore one path can be continuously transformed into the other.

Let ${\omega_1,\,...,\,\omega_n}$ be smooth differential 1-forms on ${M}$ and let us write ${\gamma ^{\star} (\omega_i)=f_i(t)dt}$ for the pull-back of each 
1-form to the interval ${[0,\,1]}$. The \emph{iterated integral} of ${\omega_1,\,...,\,\omega_n}$ along ${\gamma}$ is defined by
\begin{eqnarray}
\int_{\gamma} \omega_n ...\omega_1 = \int_{0\leq t_1 \leq ... \leq t_n \leq 1} f_n (t_n) dt_n ... f_1 (t_1) dt_1. \label{eq: iterated integral}
\end{eqnarray}
We will use the term iterated integral for ${k}$-linear combinations of such integrals.

To give an example, the multiple polylogarithms in one variable can be written as
\begin{eqnarray}
\textrm{Li}_{n_1 ,\, ...,\, n_r}(z) = (-1)^r \int_{\gamma}  \omega_0 ^{n_r-1} \omega_1 \ldots \omega_0 ^{n_1 - 1}  \omega_1 \label{eq: multiple polylog} 
\end{eqnarray}
where ${\gamma}$ is a smooth path in ${\mathbb{C} \backslash \left\{0,\,1 \right\} }$ with endpoint ${\gamma (1)=z}$ and  the sequence of 1-forms is built up from the set
 $\tilde{\Omega}_1=\left\{\omega_0 ,\, \omega_1 \right\}$ 
with ${\omega_0=\frac{dt}{t},\,  \omega_1=\frac{dt}{t-1}.}$ Note that if we extend the set ${\tilde{\Omega}_1}$ only by 1-forms of the form ${f\,dt}$, where $f$ is a rational function, possibly involving further parameters, 
and use it to define a new class of iterated integrals in a similar way, then these functions will still be defined on a one-dimensional space, corresponding to the 
one endpoint variable ${z}$. In this case let us speak of \emph{iterated integrals in one variable}.  The manifold  $M$ is an open subset of $\mathbb{C}$.
In the next section, $M$ is a certain open subset of $\mathbb{C}^n$, and we use 1-forms with several ${dt_1,\,...,\, dt_n}$   to construct a class of functions in ${n}$ variables ${z_1,\,...,\,z_n}$, given by the coordinates of the endpoint of a path.

The iterated integral in eq.\ \ref{eq: multiple polylog} is a meaningful expression for multiple polylogarithms because it depends locally on the endpoint variable ${z}$ but not on 
the path ${\gamma}$. In fact if in eq.\ \ref{eq: multiple polylog} we replace ${\gamma}$ by another path homotopic to ${\gamma}$ then we obtain the same function. 
This property is called \emph{homotopy invariance}. In the case of one-fold integrals of a 1-form ${\omega}$ one can show that homotopy invariance, 
\begin{eqnarray}
 \int_{\gamma_1} \omega = \int_{\gamma_2} \omega  \textrm{ for } \gamma_1, \, \gamma_2 \textrm{ homotopic,} \label{eq: invariant}
\end{eqnarray}
 is true, if and only if $\omega$ is closed.

For iterated integrals the condition is more complicated. It was studied in a very general setting in
 Chen's foundational work on iterated integrals \cite{Che} and we want to briefly rephrase the statement which is relevant in our context. To this end we consider tensor products of differential 1-forms
${\omega_1 \otimes ... \otimes \omega_m}$ over some field $K\subseteq k$  (which will typically be the field of rationals $\mathbb{Q}$ in the sequel) for which we use the customary bar notation ${\left[ \omega_1 | ... | \omega_m \right] }$.

 Let ${\Omega}$ be a finite set of smooth 1-forms on ${M}$ and let ${D}$ be the $K$-linear  map from tensor products of such 1-forms to tensor products of all forms on $M$, defined by
\begin{eqnarray}
D\left( \left[ \omega_1 | ... | \omega_m \right] \right) = \sum_{i=1} ^m \left[ \omega_1 | ... |\omega_{i-1} |d\omega_i | \omega_{i+1}|...| \omega_m \right] 
+ \sum_{i=1} ^{m-1} \left[ \omega_1 | ... |\omega_{i-1} |\omega_i \wedge \omega_{i+1} |...| \omega_m \right].
\end{eqnarray}
We furthermore define 
\begin{eqnarray}
B_m(\Omega) = \left\{\xi=\sum_{l=0} ^m \sum_{i_1 ,\,...,\, i_l}  c_{i_1,\ldots, i_l} \left[ \omega_{i_1} | ... | \omega_{i_l} \right] \textrm{ with }  c_{i_1,\ldots, i_l} \in K \hbox{ and }\omega_i \in \Omega \textrm{ such that } D\xi=0\right\}\ ,
\end{eqnarray}
which is a vector space over $K$.
We call the elements of this vector space \emph{integrable words}  (or \emph{bar elements})  in ${\Omega}$ and the equation ${D\xi=0}$ is known as the \emph{integrability condition}.
Now on the elements of ${B_m(\Omega)}$ let us consider the integration map, defined by simply integrating over the 1-forms according to definition  (\ref{eq: iterated integral}):
\begin{eqnarray}
\sum_{l=0}^m \sum_{i_1 ,\,...,\, i_l}  c_{i_1,\ldots, i_l} \left[ \omega_{i_1} | ... | \omega_{i_l} \right] & \mapsto & \sum_{l=0}^m \sum_{i_1 ,\,...,\, i_l}  c_{i_1,\ldots, i_l} \int_{\gamma} \omega_{i_1}...\omega_{i_l}  \label{eq: integration map} 
\end{eqnarray}

Chen's theorem now states, under some conditions on $\Omega$ which will always be satisfied in the sequel,  that this integration map gives an isomorphism from ${B_m(\Omega)}$ to the set of homotopy invariant iterated integrals in the set of 1-forms in ${\Omega}$ of length less than or equal to ${m}$.

In other words, when we apply the integration map to an integrable word, we obtain a homotopy invariant integral. The reverse is also true: any linear combination of tensor products, 
corresponding to a homotopy invariant iterated integral in the above sense, is an integrable word. In the following we implicitly use this isomorphism and represent a homotopy invariant iterated
integral by its bar element (this requires fixing basepoints: we shall demand that the regularised value of all functions at the origin is zero).
Let us consider all integrable words of a given alphabet ${\Omega}$ up to length ${m}$,  $B_m(\Omega)$ and define ${\mathcal{B}_m\left( \Omega \right)}$
to be the $K $-vector space of all the corresponding homotopy invariant iterated integrals, obtained from these words via the integration map of eq.\ \ref{eq: integration map}. 
For notational convenience, the bar notation, used for words in $B_m(\Omega)$ above, will from here on denote the corresponding functions in ${\mathcal{B}_m\left( \Omega \right)}$ as well. 
In the following section we explicitly construct this vector-space of functions for a specific choice of ${\Omega}$.

\section{Universal multiple polylogarithms in several variables \label{sec: polylogs}}

From now on, let $K= \mathbb{Q}$. Extending the set ${\tilde{\Omega}_1}$ of the previous section, let us define the auxiliary set of differential 1-forms
\begin{eqnarray}
 \tilde{\Omega}_n = \left\{ \frac {dt_1}{t_1}, \frac {dt_1}{t_1-1}, \frac {t_2 dt_1}{t_1 t_2 -1},\, ...,\, \frac {  \left( \Pi_{i=2} ^n t_i \right)  dt_1 }{ \Pi_{i=1} ^n t_i  -1} \right\} \label{eq: Omega1}
\end{eqnarray}
on an open subset of ${\mathbb{C}}$ with coordinate ${t_1}$. As the 1-forms in ${\tilde{\Omega}_n}$ are closed and furthermore the wedge product of each pair 
of 1-forms in  ${\tilde{\Omega}_n}$ is zero, the integrability condition is trivially satisfied for any word in the  letters of ${\tilde{\Omega}_n}$, and therefore ${B_m \left( \tilde{\Omega}_n \right)}$ is spanned by the set of all words of length $\leq {m}$ in ${\tilde{\Omega}_n}$. Therefore it is trivial to obtain the 
homotopy invariant iterated integrals ${\mathcal{B}_m \left( \tilde{\Omega}_n \right).}$

However, all of these are homotopy invariant functions of only one variable, the endpoint of the path corresponding to the integration over ${dt_1}$. By a slight abuse of notation let us call this variable ${t_1}$. The other parameters 
${t_2,\, ...,\,t_n}$ in the 1-forms have to be treated as constants up to now.

Let us now consider the set
\begin{eqnarray}
 \Omega_n = \left\{ \frac {dt_1}{t_1},...,\, \frac {dt_n}{t_n}, \frac{d \left( \Pi_{a\leq i \leq b} t_i \right) }{ \Pi_{a \leq i \leq b} t_i  -1}  \textrm{ where } 1\leq a\leq b \leq n \right\}. \label{eq: Omega}
\end{eqnarray}
For example, in the case of three variables we have
\begin{eqnarray}
 \Omega_3 =  \left\{ \frac {dt_1}{t_1},\,\frac {dt_2}{t_2},\, \frac {dt_3}{t_3}, \frac{t_1 dt_2 + t_2 dt_1}{t_1 t_2 -1},\, \frac{t_2 dt_3 + t_3 dt_2}{t_2 t_3 -1}, \, \frac{t_1 t_2 dt_3 + t_2 t_3 dt_1 + t_1 t_3 dt_2}{t_1 t_2 t_3-1} \right\}.\, \, 
\end{eqnarray}
Note that for ${n>1}$, in contrast to the previous case, not every possible word in ${\Omega_n}$ belongs to ${B_m \left( \Omega_n \right)}$. We explicitly construct ${\mathcal{B}_m \left( \Omega_n \right)}$ by a map 
\begin{eqnarray}
 \psi: \mathcal{B}_m \left( \tilde{\Omega}_n \right) \rightarrow \mathcal{B}_m \left( \Omega_n \right),
\end{eqnarray} 
which is defined as follows.

Let $F_n$ be the vector space of rational functions of ${t_1,\,...,\,t_n}$ with coefficients in $\mathbb{Q}$ whose denominators are 
products of elements in the set $\{t_1,\ldots, t_n, \prod_{a \leq i \leq b} t_i-1\}$, for $1\leq a \leq b \leq n$.
In the following we write iterated integrals in ${\mathcal{B}_m \left( \tilde{\Omega}_n \right)}$ as ${\left[ g_1 dt_1 |  g_2 dt_1 | ... | g_n dt_1 \right]}$, where $g_i \in F_n$.
Differentiation of the iterated integrals in ${\mathcal{B}_m \left( \tilde{\Omega}_n \right)}$ with respect to ${t_1}$ can be computed by
\begin{equation} \label{dt1}
{\frac{\partial}{\partial t_1} \left[ g_1 dt_1 \right]  =  g_1 \,\,  \textrm{ and }  \,\,\frac{\partial}{\partial t_1} \left[ g_1 dt_1 |  g_2 dt_1 | ... | g_n dt_1 \right]  =   g_1 \left[g_2 dt_1 | ... | g_n dt_1 \right] \textrm{ for } n\geq 2.}
\end{equation}
The map ${\psi}$ will not change the differential behaviour with respect to ${t_1}$, so we impose the differential equations 
$${\frac{\partial}{\partial t_1} \psi \left( \left[ g_1 dt_1 \right] \right)  =  g_1  \,\, \textrm{ and } \,\, \frac{\partial}{\partial t_1} \psi \left( \left[ g_1 dt_1 |  g_2 dt_1 | ... | g_n dt_1 \right] \right)  =   g_1 \psi \left( \left[g_2 dt_1 | ... | g_n dt_1 \right] \right) \textrm{ for } n\geq 2.}$$

Differentiation with respect to ${t_2}$ is not defined on ${\xi \in \mathcal{B}_m \left( \tilde{\Omega}_n \right)}$, but we want ${\psi(\xi)}$ to have a well-defined differential behaviour with respect to ${t_2}$. 
To this end, we consider an auxiliary operator 
\begin{eqnarray}
\partial_{t_2}: \mathcal{B}_m \left( \tilde{\Omega}_n \right) \rightarrow F_2 \otimes \mathcal{B}_m \left( \tilde{\Omega}_n \right) 
\end{eqnarray}
where 
${\partial_{t_2}}$ is defined by the following properties: \\
(a) On rational functions it acts as differentiation with respect to ${t_2}$: ${\partial_{t_2}g = \frac{\partial}{\partial t_2}g}$. \\
(b) It commutes with differentiation with respect to ${t_1}$: ${\partial_{t_2} \frac{\partial}{\partial t_1} \xi = \frac{\partial}{\partial t_1} \partial_{t_2} \xi}$.\\
Using property (b) we obtain 
\begin{eqnarray}
\frac{\partial}{\partial t_1} \partial_{t_2}  \left[ g_1 dt_1 | g_2 dt_1 | ... | g_n dt_1 \right]  & = &  \partial_{t_2} \frac{\partial}{\partial t_1}  \left[ g_1 dt_1 | g_2 dt_1| ... | g_n dt_1 \right] \\
 & = & \partial_{t_2} g_1  \left[ g_2 dt_1 | ... | g_n dt_1 \right] 
\end{eqnarray}
and therefore 
\begin{eqnarray} \label{dt2asint}
\partial_{t_2}  \left[ g_1 dt_1 | g_2 dt_1  |... | g_n dt_1 \right]  & = & \int_0 ^{t_1} dt'_1 \partial_{t_2} g_1  \left[ g_2 dt'_1  |... | g_n dt'_1 \right] .
\end{eqnarray}
Note that on the right hand side of the last equation, ${\partial_{t_2}}$ acts on an iterated integral of length ${n-1}$, so we have a recursive procedure to compute ${\partial_{t_2} \xi}$, 
the last stage of the recursion given by property (a).   In   $(\ref{dt2asint})$, the integral is computed by decomposing $\partial_{t_2} g_1\in F_n$ into partial fractions with respect to $t_1'$, and using integration by parts and the formula 
$(\ref{dt1})$.

Now we define ${\psi}$ such that it satisfies 
$${\frac{\partial}{\partial t_2} \psi (\xi) =  \psi (\partial_{t_2} \xi)}$$
and analogous differential equations with respect to the remaining parameters ${t_i}$. 
The constants are fixed by demanding that 
 ${\psi (\xi)}$ has a finite  expansion at the origin of the form
\begin{equation}
\sum_{0 \leq i_1, \ldots,  i_n \leq N} f_{i_1,\ldots, i_n} (\log  t_1)^{i_1} \ldots (\log t_n)^{i_n}
\end{equation}
  where $f_{i_1,\ldots, i_n}$ is analytic at the origin  and vanishes at the point ${t_1=...=t_n=0}$. 
Together with the above differential equations this determines the map ${\psi}$. This map is closely related to constructions which were recently introduced to the physics literature as the `symbol' \cite{Gon09, Gon10, Duh11}.

To give an example, we apply ${\psi}$ to
\begin{eqnarray}
\xi  =  \left[ \frac{dt_1}{t_1 -1} \left| \frac{t_2 \,dt_1}{t_1 t_2-1} \right.  \right] \in \mathcal{B}_2(\Omega_2).
\end{eqnarray}
We obtain
\begin{eqnarray}
\frac{\partial}{\partial t_1} \psi (\xi) & = & \frac{1}{t_1-1} \left[ \frac{t_1 \, dt_2 + t_2 \, dt_1}{t_1 t_2-1} \right],\\
\frac{\partial}{\partial t_2} \psi (\xi) & = & \frac{1}{t_2-1} \left[ \frac{dt_1}{t_1-1} \right] + \left( \frac{1}{t_2} - \frac{1}{t_2-1} \right)  \left[ \frac{t_1 \, dt_2 + t_2 \, dt_1}{t_1 t_2-1} \right],\\
\psi (\xi) & = & \left[ \frac{dt_2}{t_2-1} \left| \frac{dt_1}{t_1-1} \right. \right] + \left[\frac{dt_1}{t_1-1} + \frac{dt_2}{t_2} - \frac{dt_2}{t_2-1} \left| \frac{t_1 \, dt_2 + t_2 \, dt_1}{t_1 t_2-1} \right. \right].
\end{eqnarray}

This iterated integral ${\psi (\xi)}$ is in fact equal to the multiple polylogarithm in two variables ${\textrm{Li}_{1,\,1}(t_1,\,t_2)}$ while the 
expression $\xi$ coincides with this function only on a one-dimensional subspace for fixed ${t_2}$ and does not capture its differential behaviour with respect to
${t_2}$.

Let  ${\mathcal{B} \left( \Omega_n \right)}=\sum_{m \geq  0} {\mathcal{B}_m \left( \Omega_n \right)} $ denote the vector space of integrable words of all lengths $m\geq 0$. It 
 was extensively studied by one of us in the context of the moduli space of curves of genus zero with ${m+3}$ marked points, 
${\mathcal{M}_{0,\,m+3},}$
in reference \cite{Bro06} and we refer to this work for details and proofs of the following properties:

\begin{itemize}

\item The elements of ${\mathcal{B} \left( \Omega_n \right)}$ are homotopy invariant and therefore they are functions of ${n}$ variables.
\item ${\mathcal{B}\left( \Omega_n \right)}$ contains the multiple polylogarithms of Goncharov.
\item There is an explicit basis for ${\mathcal{B} \left( \Omega_n \right)}$ in terms of the map $\Psi$.   There is a decomposition  
 $\mathcal{B} \left( \Omega_n \right) = \Psi( \mathcal{B}\left( \tilde{\Omega}_n \right)) \otimes \ldots \otimes \Psi( \mathcal{B}\left( \tilde{\Omega}_2 \right)) \otimes \Psi( \mathcal{B}\left( \tilde{\Omega}_1 \right))$ and we have an explicit basis for each 
${\mathcal{B}\left( \tilde{\Omega}_i \right)}$ given by the set of words in $\tilde{\Omega}_i $, by the discussion after  (\ref{eq: Omega1}).
\item ${\mathcal{B} \left( \Omega_n \right)}$ is closed under taking primitives.
\item The limits of elements of ${\mathcal{B} \left( \Omega_n \right)}$ at $t_n$ equal to 0 and 1 are ${\mathcal{Z}}$-linear combinations of elements of ${\mathcal{B} \left( \Omega_{n-1} \right)}$, where 
${\mathcal{Z}}$ is the ${\mathbb{Q}}$-vector space of multiple zeta values.
\end{itemize}

As a consequence of the latter properties, we can evaluate definite integrals of the type
\begin{eqnarray}
\int_0 ^1 dt_n \sum _j f_j \beta_j , \quad \,\beta_j\in \mathcal{B}_m \left( \Omega_n \right) ,  \quad f_j \in F_n. \label{eq: integrable}
\end{eqnarray}
The result  will be  a
${\mathcal{Z}}$-linear combination  of elements of ${\mathcal{B}_{m} \left( \Omega_{n-1} \right)}$ multiplied by  elements of ${F_{n-1}}$. This can be iterated. 
Our forthcoming publication will feature algorithms and a computer program for the computation of such integrals.

\section{Application to Feynman parametric integrals \label{sec: Feynman}}

The integrals $(\ref{eq: integrable})$ play a role in pure mathematics, such as in reference \cite{Bro06, BroCarSch}, and in physics in the context of deformation quantization \cite{FW}, 
superstring theory \cite{SchSti}, Schnetz' model of graphical functions \cite{Sch}
and in perturbative quantum field theory. Here we focus on the latter and give a very brief outlook on how the use of ${\mathcal{B}_m \left( \tilde{\Omega}_n \right)}$ can facilitate the computation of 
Feynman integrals. We follow the approach of reference \cite{Bro08} which in combination with the use of hyperlogarithms already led to new results for certain integrals relevant in QCD \cite{DESY-Z}. 
Other results were recently obtained by similar strategies of integrating over Feynman parameters, e.g.  in \cite{ChaDuh}.

To begin with, we consider a primitive (subdivergence-free) overall logarithmically divergent vacuum Feynman graph, giving rise to a finite integral 
\begin{eqnarray}
I & = & \int_0 ^\infty ... \int_0 ^\infty \left( \Pi_{i=1} ^N dx_i \right) \delta \left( 1-\sum_{i=1} ^N x_i \right) \frac{1}{\mathcal{U}^2}
\end{eqnarray}
over ${N}$ Feynman parameters with ${\mathcal{U}}$ being the first Symanzik polynomial (see e.g.\ \cite{Nak, ItzZub}). Reference \cite{Bro08} provides a polynomial reduction algorithm which for any ordered sequence
 of the ${N}$ Feynman parameters  ${\lambda=\left( x_{\sigma_1},\,...,\,x_{\sigma_N} \right)}$ gives a sequence ${\mathcal{S}_\lambda = \left( S_1 ,\, ... ,\, S_N \right)}$ of sets of
polynomials in the Feynman parameters. Without repeating the details of this algorithm here, we recall that one can evaluate ${I}$ by iteratively integrating over the Feynman 
parameters in the order ${\lambda}$ if in each set ${S_i\in \mathcal{S}_\lambda}$ all polynomials are linear in the corresponding parameter ${x_{\sigma_i}}$.

Note that the latter is exactly the condition for the integral to be computable by our program, using the functions  in $\mathcal{B}_m$. Indeed, let ${\lambda}$ be a sequence for which this criterion holds
and assume that we already integrated out the first ${i-1}$ parameters. After mapping the integration domain of ${dx_{\sigma_i}}$ to ${[0,\,1]}$ we may assume that the integrand 
is of the form 
\begin{eqnarray}
I_i & = &  \sum_m \sum _j f_j \left[ \omega_{j,\,1} | ... | \omega_{j,\,m} \right],
\end{eqnarray}
where ${f_j}$ are algebraic functions and ${\left[ \omega_{j,\,1} | ... | \omega_{j,\,m} \right]\in \mathcal{B}_m ( \Omega_n )}$ for some ${m}$ and ${n}$. 
The denominators of ${f_j,\,\omega_{j,\,1},\,...,\,\omega_{j,\,m}}$ are irreducible polynomials which map to the 
members of ${S_i}$. If these polynomials are linear in ${x_{\sigma_i}}$ then there is a ${k}$ such that we can map these polynomials to the denominators of the 1-forms 
in ${\Omega_k}$ of eq.\ \ref{eq: Omega} and we can express ${I_i}$ by terms of the form of eq. \ref{eq: integrable}.

It is an important advantage of this approach that by the polynomial reduction algorithm, i.e. by simple operations on polynomials and without integrating, 
we can decide whether the method applies and which order of parameters we should choose. Another advantage of the use of ${\mathcal{B}_m \left( \Omega_n \right)}$ is
that if the polynomial reduction can be done, ${\mathcal{Z}}$-linear combinations of these functions are sufficient to express the intermediate results after each integration.

%\section{Conclusions \label{sec: conclusions}}

It was shown in reference \cite{Bro09} that the method is applicable for a large class of graphs and it is well-known how to relate vacuum-graphs to contributions to two-point functions \cite{CheTka}.
For graphs with further legs and with non-zero particle masses, we have to take the second Symanzik polynomial into account. Certain properties
of this polynomial \cite{BogWei} give rise to the hope that the polynomial reduction and the above method can be extended to a large number of such Feynman graphs as well. It is
furthermore important for us to move beyond the restriction of primitive graphs. Recent work of Kreimer and one of us \cite{BroKre, BroKre2} provides a strategy to express Feynman integrals
with UV-subdivergences by integrals for which the above method can be applied. The methods of \cite{KSV} suggest that this method also generalizes to gauge theories.

\end{document}